\documentclass{llncs}
\usepackage{amsmath}
\usepackage{booktabs} 
\usepackage{caption} 
\usepackage{graphicx}
\usepackage{pgfplots}
\usepackage[all]{nowidow}
\usepackage[utf8]{inputenc}
\usepackage{tikz}
\usetikzlibrary{er,positioning,bayesnet}
\usepackage{multicol}
\usepackage{hyperref}
\usepackage[inline]{enumitem} 
\usepackage{wrapfig}
\usepackage[caption=false]{subfig}

\definecolor{blue}{HTML}{1F77B4}
\definecolor{orange}{HTML}{FF7F0E}
\definecolor{green}{HTML}{2CA02C}

\pgfplotsset{compat=1.14}

\usepackage{multirow}
\usepackage{amsmath,amsfonts,mathtools}
\usepackage{diagbox}
\usepackage{amssymb}

\usepackage{algorithm}
\usepackage[noend]{algpseudocode}

\algnewcommand\algorithmicforeach{\textbf{for each}}
\algdef{S}[FOR]{ForEach}[1]{\algorithmicforeach\ #1\ \algorithmicdo}

\usepackage[font=footnotesize]{caption}
\usepackage{subfig}
\usepackage{booktabs} 
\usepackage{multirow}
\usepackage{amsmath,amsfonts,mathtools}
\usepackage{diagbox}

\algtext*{EndWhile}
\algtext*{EndIf}
\algtext*{EndFor}

\usepackage{pifont}
\newcommand{\cmark}{\ding{51}}%
\newcommand{\xmark}{\ding{55}}%

\begin{document}

\title{On Designing Secure and Robust Scan Chain \\ for Protecting Obfuscated Logic}

\author{Hadi Mardani Kamali$^*$ \and Kimia Zamiri Azar$^*$ \and \\ Houman Homayoun$^+$ \and Avesta Sasan$^*$}

\institute{$^*$Department of Electrical and Computer Engineering, \\
George Mason University, Fairfax, VA, USA. \\
\email{\{hmardani, kzamiria, asasan\}@gmu.edu} \\
\textcolor{white}{..\\}
$^+$Electrical and Computer Engineering Department, \\
University of California Davis, Davis, CA, USA. \\
\email{hhomayoun@ucdavis.edu}}

\maketitle

\begin{abstract}

In this paper, we assess the security and testability of the state-of-the-art design-for-security (DFS) architectures in the presence of scan-chain locking/obfuscation, a group of solution that has previously proposed to restrict unauthorized access to the scan chain. We discuss the key leakage vulnerability in the recently published prior-art DFS architectures. This leakage relies on the potential glitches in the DFS architecture that could lead the adversary to make a leakage condition in the circuit. Also, we demonstrate that the state-of-the-art DFS architectures impose some substantial architectural drawbacks that moderately affect both test flow and design constraints. We propose a new DFS architecture for building a secure scan chain architecture while addressing the potential of key leakage. The proposed architecture allows the designer to perform the structural test with no limitation, enabling an untrusted foundry to utilize the scan chain for manufacturing fault testing without needing to access the scan chain. Our proposed solution poses negligible limitation/overhead on the test flow, as well as the design criteria.

\end{abstract}


The large cost of building semiconductor fabrication foundries, the ever-increasing density of ICs using smaller technology nodes, and significant maintenance and operation costs of the semiconductor facilities, have forced many high-tech companies to outsource many design stages, including the fabrication \cite{yeh2012trends}, making IC supply chain a global one. This financially and economically sensible vertical supply-chain model, however, has raised many security and trust concerns including but not limited to  IP piracy, reverse engineering, and IC overproduction \cite{rostami2014primer}. 

To combat these threats, amongst many countermeasure solutions, logic obfuscation \cite{roy2008epic} introduces a form of post-manufacturing programming into the design, making the functionality of the circuit dependent to the programming values, referred to as the \emph{key}. After fabrication, when the design house receives the fabricated ICs, the correct key is programmed into a tamper-proof non-volatile memory (tpNVM) \cite{tuyls2006read}, reducing the functionality of the IP to the correct functionality. If the inserted key is not the correct key, the obfuscated design implement a different bogus function, generating a different output response (output corruption) to the same input.


In 2015, the introduction of the Boolean satisfiability (SAT) attack \cite{subramanyan2015evaluating,el2015integrated} challenged the validity/strength of all prior logic obfuscation solutions. The SAT attack was able to easily break logic obfuscation solutions in a matter of seconds to minutes, shattering the false sense of security of all prior-art logic obfuscation solutions. The original SAT attack is applicable to combinational circuits. However, the existence of the much-needed scan chain for functional and structural testing, makes the sequential circuits also vulnerable to this attack when an adversary gains access to the scan chain.
 
After the introduction of the SAT attack, a wide range of new logic obfuscation solutions have been introduced in the literature to combat/prevent the SAT attack. These countermeasures could be divided into logic obfuscation solutions aiming to (1) formulate and apply a SAT-resilient logic obfuscation solution, and those aiming to (2) restrict unauthorized access to the scan chain. 

Regarding the former category, they could be further broken down into (a) logical SAT-hard solutions such as SFLL \cite{yasin2017provably} and Full-lock \cite{kamali2019full}, and (b) behavioural SAT-inapplicable obfuscation techniques, such as cyclic logic locking \cite{shamsi2017cyclic}, or Delay Logic Locking (DLL) \cite{xie2017delay}. Logical SAT-hard solutions aim at increasing either the number of SAT attack iterations \cite{yasin2017provably} or the number of recursive calls in each iteration of the SAT attack to a sufficiently large number \cite{kamali2019full}. However, the problem with this group is either the extremely low output corruption (in SFLL \cite{yasin2017provably}) or a large area overhead (in Full-lock \cite{kamali2019full})\nocite{kamali2018lut}. On the other hand, the behavioral SAT-inapplicable obfuscation techniques aim to build techniques in a way that cannot be modeled using the SAT attack, such as cycles that trap the SAT solver or delay locking that cannot be modeled by the SAT attack. However, this breed was later broken by SAT inspired attacks on cyclic obfuscation such as CycSAT \cite{zhou2017cycsat}\nocite{roshanisefat2018srclock} and satisfiability modulo theory (SMT) attack \cite{azar2019smt,zamiri2019threats}. 

Blocking unauthorized access to the scan chain as the latter category limits the access of an adversary only to the primary inputs and primary outputs (PI/PO) \cite{karmakar2018encrypt,wang2017secure,guin2018robust,wang2019secure,azar2019coma}. Expanding on the SAT attack, it was later shown, that an adversary can still attack a sequential circuit with no access to the scan chain by using an unrolling-based SAT attack \cite{el2017reverse} or a bounded-model-checking (BMC) attack \cite{alrahis2019scansat}. However, these sequential attacks are far weaker than pure SAT and are mostly applicable to moderately small sequential circuits. Since the sequential attacks are not scalable, by blocking the scan chain, and applying many of the prior logic obfuscation techniques, a moderately size obfuscated netlist could easily resist such attacks.   

Prior work on restricting unauthorized access to the scan chain could be divided into (1) scan chain obfuscation \cite{karmakar2018encrypt,wang2017secure} and (2) scan chain blocking \cite{guin2018robust,wang2019secure}. In the scan chain obfuscation techniques, such as encrypt flip-flop \cite{karmakar2018encrypt} or dynamically obfuscated scan (DOS) \cite{wang2017secure}, the scan chain is statically or dynamically locked by inserting key gates. However, ScanSAT \cite{alrahis2019scansat} could break both statically and dynamically scan chain obfuscation techniques by transforming the obfuscated scan chain into a combinational circuit and thereby launching the SAT attack on them (the unrolling-based SAT) \cite{alrahis2019scansat}. 

In the scan chain blocking techniques, after loading the obfuscation key (from tpNVM), the access to the scan-out(s) (SO) would be blocked \cite{guin2018robust}. By eliminating the access to the SO, an adversary's ability to monitor the behavior of the circuit will be limited only to the PO. This eliminates the possibility of the SAT attack as well as any attack that requires access to the scan chain, forcing an attacker to use the far weaker and non-scalable sequential attacks. 

Scan chain blocking in the presence of logic obfuscation was first introduced in \cite{guin2018robust}.  In the rest of this paper we refer to this solution as \textbf{r}obust \textbf{d}esign-\textbf{f}or-\textbf{s}ecurity (R-DFS). In addition to blocking the SO, the R-DFS also introduces a new storage element for holding the obfuscation key, denoted by secure cell (SC). However, the security of the R-DFS architecture was later challenged by the \emph{shift-and-leak} attack \cite{limaye2019robust}. To remedy the leakage issue, the authors proposed modification to the scan blocking architecture (we call it \textbf{mR-DFS}), equipping the SCs with a mode switch shift disable (MSSD) circuitry \cite{limaye2019robust}. The mR-DFS blocks any shift operation after the obfuscation key is loaded from the tpNVM, removing the ability of an adversary to apply the shift-and-leak attack. 



In this paper, by showing the architectural drawbacks of mR-DFS, we introduce our proposed DFS scan blocking architecture for protecting the logic obfuscation key. More precisely, the contributions of this work are as follows: (1) We illustrate how a glitch-based shift-and-leak attack allows an adversary to leak the logic obfuscation key even if the shift operation is disabled in mR-DFS, thereby, leaking the actual logic obfuscation key through the PO. (2) As a countermeasure, we propose a new key-trapped design-for-security (\textbf{kt-DFS}) architecture, where the scan chain that loads the logic obfuscation key is fully detached from regular scan chain(s). To fulfill this requirement, we propose a new secure cell design content of which cannot be shifted in the scan chain after a key registration event is observed. (3) We assess the security of proposed kt-DFS, and compare the proposed solution with R-DFS and mR-DFS. As shown in Table \ref{compare}, we will illustrate how the kt-DFS can support both structural and functional testing while resisting all leaky-based and SAT-based attacks on logic locking.

\begin{table}
\footnotesize
\centering
\caption{Comparison of the Existing DFS architectures with our proposed kt-DFS.}
\label{compare}
\setlength\tabcolsep{2pt}
\begin{tabular}{@{} lccccc @{}}
\toprule 
\multirow{2}{*}{Defenses} & Test & Test  & \multicolumn{3}{c}{Resilient against} \\
& Time & Complexity & ScanSAT \cite{alrahis2019scansat} & Shift\&leak \cite{limaye2019robust} & Glitch\&Leak \\
\cmidrule(r){1-1} \cmidrule(lr){2-2} \cmidrule(lr){3-3} \cmidrule(lr){4-6}
EFF + RLL \cite{karmakar2018encrypt} & low & None & \textcolor{red}{\xmark} & \cmark & \cmark \\
\cmidrule(r){1-1} \cmidrule(lr){2-2} \cmidrule(lr){3-3} \cmidrule(lr){4-6}
R-DFS + SLL \cite{guin2018robust} & low & None & \cmark & \textcolor{red}{\xmark} & \cmark \\
\cmidrule(r){1-1} \cmidrule(lr){2-2} \cmidrule(lr){3-3} \cmidrule(lr){4-6}
\multirow{2}{*}{mR-DFS + SLL \cite{limaye2019robust}} & \multirow{2}{*}{\textcolor{red}{high}} & \textcolor{red}{key reload} & \multirow{2}{*}{\cmark} & \multirow{2}{*}{\cmark} & \multirow{2}{*}{\textcolor{red}{\xmark}} \\
 & & \textcolor{red}{per pattern} & \\
\cmidrule(r){1-1} \cmidrule(lr){2-2} \cmidrule(lr){3-3} \cmidrule(lr){4-6}
kt-DFS + SLL & low & None & \cmark & \cmark & \cmark \\
\bottomrule
\end{tabular}
\end{table}

\section{Background}
\label{sec:Background}

Both R-DFS \cite{guin2018robust} and mR-DFS \cite{limaye2019robust} 
block the SO pins after the obfuscation key is loaded into the design. The mR-DFS is built on top of R-DFS to fix the leakage issue. In the following section, we first describe how R-DFS works. Then, we explain the leakage issue identified in R-DFS, motivating the shift-and-leak attack. Then, we describe how mR-DFS remedies the problem with disabling shift operations after loading the obfuscation key.  

\subsection{R-DFS: Restricting Scan Access}

\begin{figure}
    \centering
    \subfloat[]{{\includegraphics[width=0.45\columnwidth]{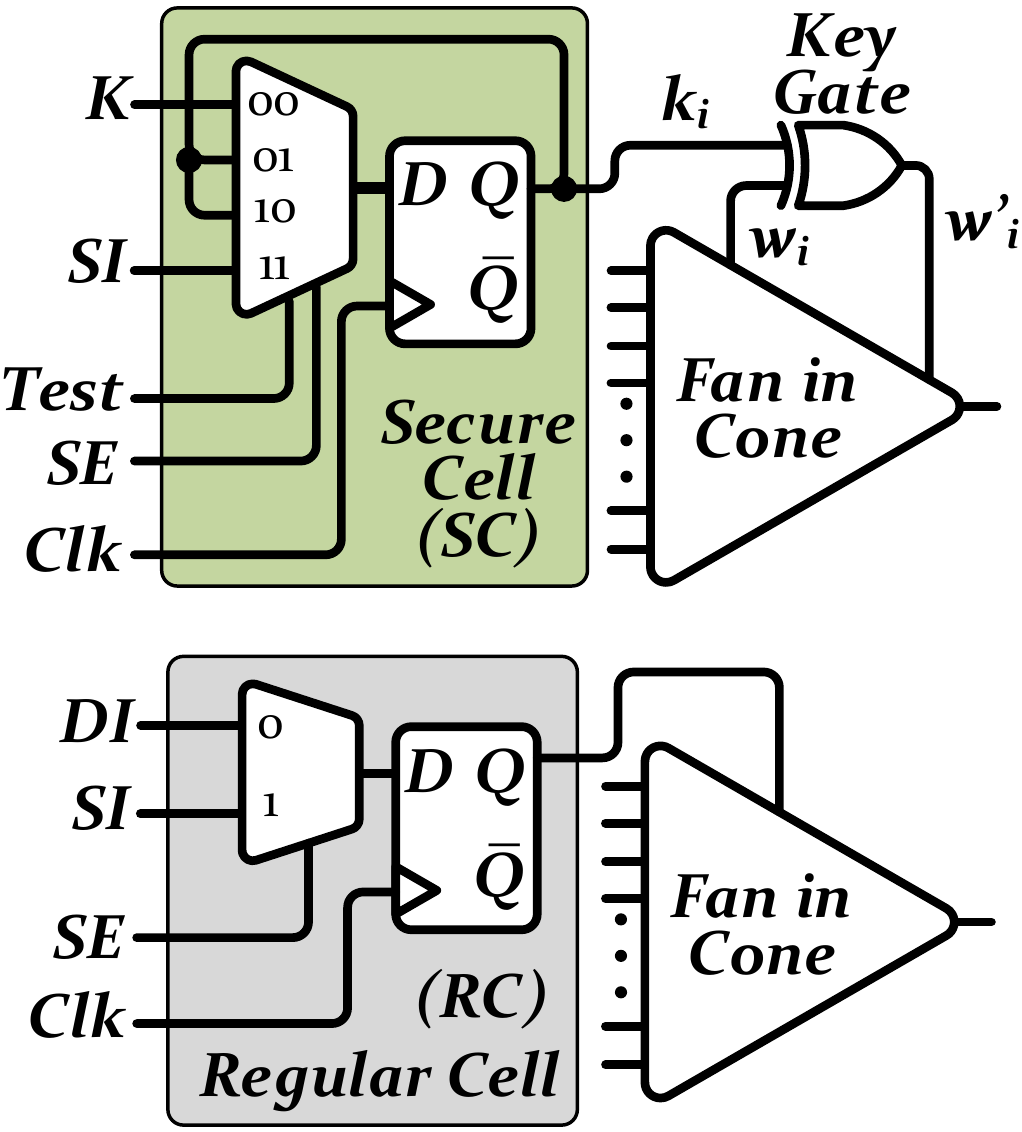}}}
    \subfloat[]{{\includegraphics[width=0.55\columnwidth]{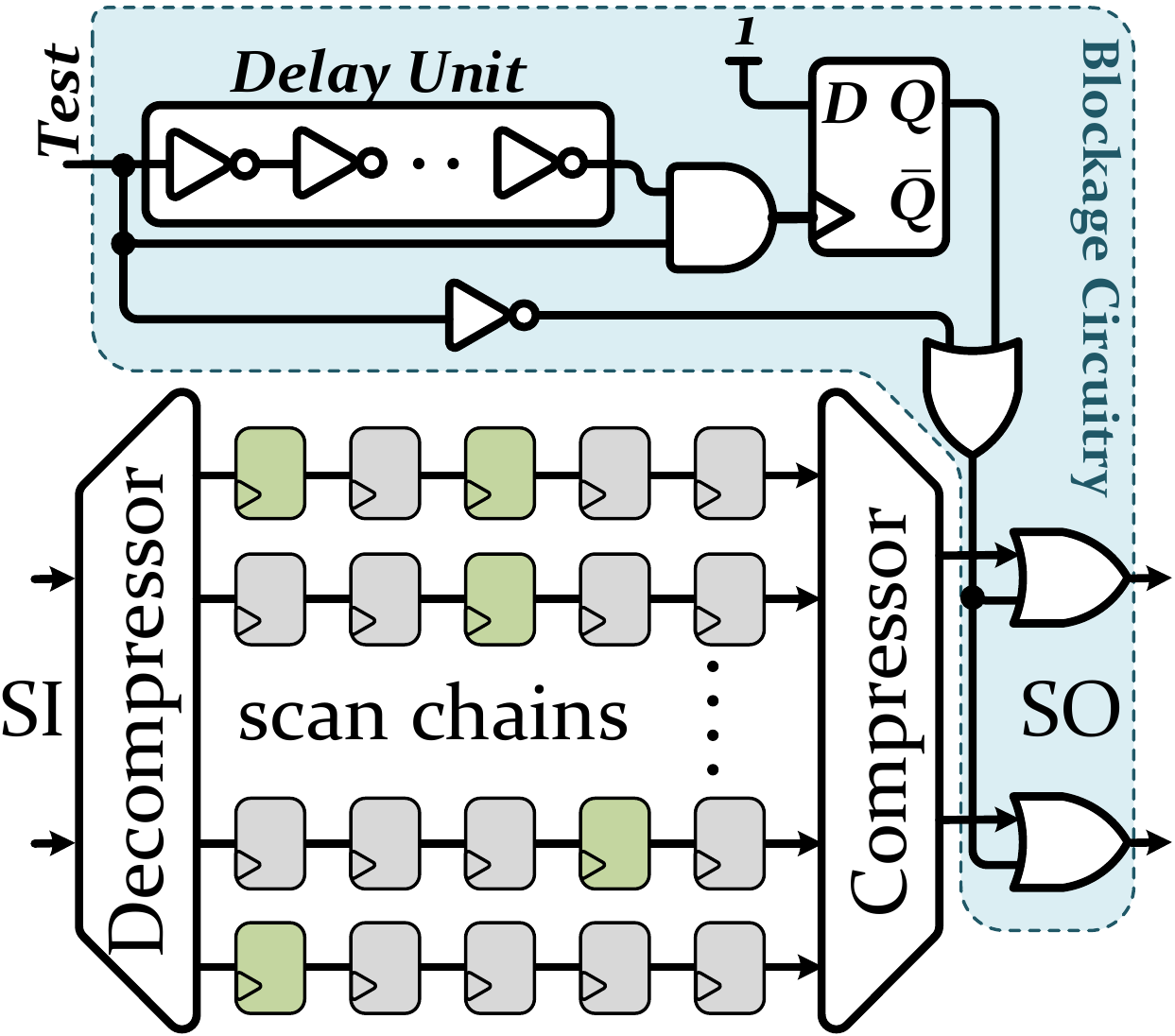}}}
    \caption{(a) Secure Cell (SC) vs. Regular Cell (RC), (b) Restricted Unauthorized Scan Access using Blockage Circuitry.}
    \label{r-dfs}
\end{figure}

In R-DFS \cite{guin2018robust}, the obfuscation key is stored in a custom-designed scan (storage) cell, denoted as secure cell (SC). As shown in Fig. \ref{r-dfs}(a), in R-DFS, each key value is stored in one SC. The R-DFS architecture, as indicated in Table \ref{mode}, allows four types/modes of operation based on the $Test$ and $SE$ pins. The key values could be loaded into SCs either directly from tpNVM (actual key values in mode $M_0$) or the scan-in (dummy/actual key in mode $M_2$). The scan chains, as shown in Fig. \ref{r-dfs}(b), are constructed by stitching the SCs with regular scan Flip-Flops (SFF). The SFFs in this paper are denoted as Regular Cells (RC). The SCs keep their previous values in modes $M_{1a}$ and $M_{1b}$. The only difference between the $M_{1a}$ and $M_{1b}$ mode is the value of the $SE$ pin that determines the shift/capture mode in RCs. Both of the $M_{1a}$ and $M_{1b}$ modes allow the SCs to be bypassed (keeping their previous values) when the RCs are in shift/capture mode. 

For the structural (a.k.a manufacturing fault) test, the $Test$ pin must be $1$, allowing the shift and capture operations to be carried in modes $M_2$ and $M_{1b}$ respectively, giving unrestricted access to the scan. On the other hand, for a functional test, first, the correct key is loaded from tpNVM into SCs using the mode $M_0$. Then, the initial state is loaded into the RCs in mode $M_{1a}$, with no change on the key value in SCs. Finally, the response is observed at the PO in mode $M_0$. To block unauthorized access to the scan chain (when a valid key is loaded), as illustrated in Fig. \ref{r-dfs}(b), the R-DFS architecture utilizes a SO-blockage circuitry. This module blocks/masks the SOs upon a switch from functional mode (mode $M_0$ that loads the actual key into SCs) to test mode (mode $M_2$ that supports the shift operation). Hence, after loading the key in mode $M_0$, SO will no longer be accessible, removing the possibility of SAT attack, and limiting the adversary's attack option to the far weaker and non-scalable unrolling-based or BMC based attacks.  

\begin{table}
\footnotesize
\centering
\caption{Modes of Operation in Secure Cell (SC).}
\label{mode}
\setlength\tabcolsep{5pt} 
\begin{tabular}{@{} cccl @{}}
\toprule 
Test & SE & Mode & Description \\
\cmidrule(lr){1-2} \cmidrule(lr){3-3} \cmidrule(lr){4-4}
\multirow{2}{*}{0} & \multirow{2}{*}{0} & \multirow{2}{*}{$M_0$} & The circuit is in functional mode. Actual keys from\\
 & & & tpNVM applies to the Logic (Correct Functionality). \\
 \cmidrule(lr){1-2} \cmidrule(lr){3-3} \cmidrule(lr){4-4}
0 & 1 & $M_{1a}$ & The SCs hold their previous value. Based on the value \\
1 & 0 & $M_{1b}$ & of SE, RCs are in capture/shift mode. \\
\cmidrule(lr){1-2} \cmidrule(lr){3-3} \cmidrule(lr){4-4}
\multirow{2}{*}{1} & \multirow{2}{*}{1} & \multirow{2}{*}{$M_2$} & The SCs become part of the scan chain. Actual/Dummy\\
 & & &  keys from SI for structural testing. \\
\bottomrule
\end{tabular}
\end{table}

\subsection{Shift-and-Leak Attack on R-DFS}

\begin{figure}
    \centering
    \subfloat[Determining the Leaky Cell (LC) in Circuit.] {{\includegraphics[width=\columnwidth]{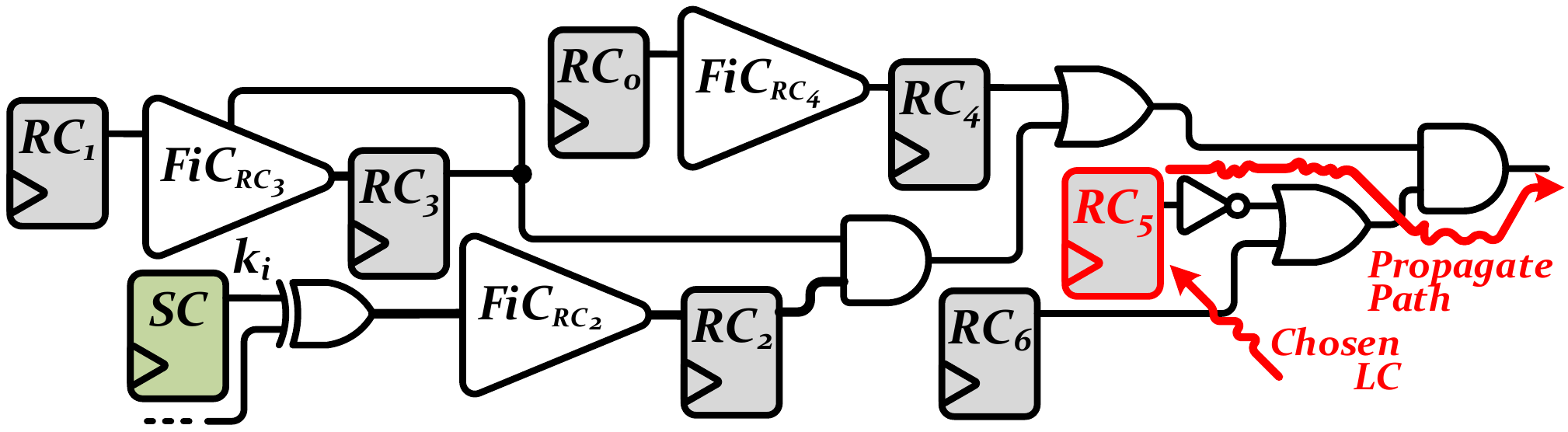}}} \\
    \subfloat[The Scan Chain Configuration when $SE$=1. ($d$ = 2)] {{\includegraphics[width=\columnwidth]{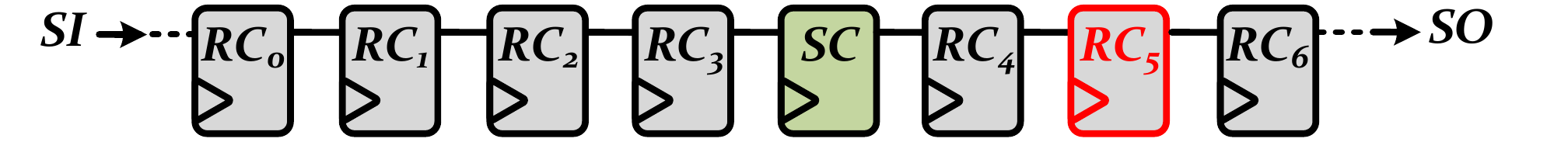}}} \\
    \subfloat[Shift-in the leaky condition ($d$-bit reverse-shifted) based on $d$=2.] {{\includegraphics[width=\columnwidth]{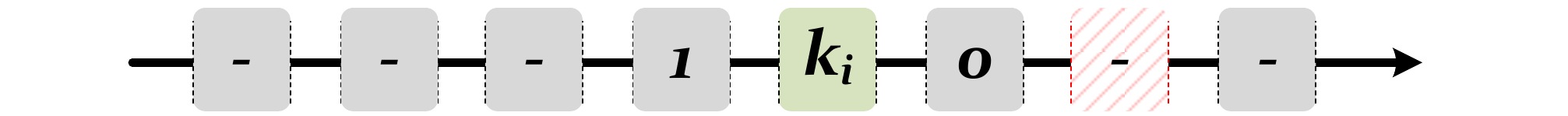}}} \\
    \subfloat[shift all FFs, including SCs and RCs, in Mode $M_2$, to put the $k_i$ into the LC.] {{\includegraphics[width=\columnwidth]{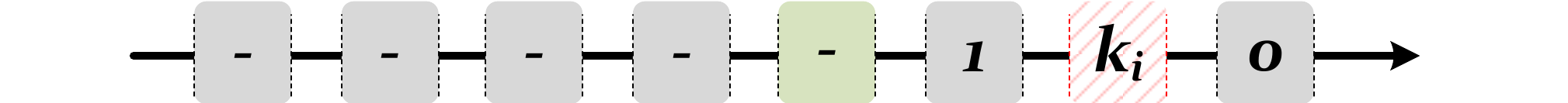}}} \\
    \caption{Example of shift-and-leak attack on R-DFS.}
    \label{shift_attack}
\end{figure}

Although R-DFS breaks the SAT attack by blocking the SO, the introduction of shift-and-leak attack \cite{limaye2019robust} shows that there is a valid key leakage possibility in R-DFS that allows the adversary to observe and extract the logic obfuscation key using PO. This attack exploits (1) the availability of the shift-in process through $SI$, and (2) the capability of reading out the PO through chip pin-outs in the functional mode. Considering Fig. \ref{shift_attack} as an illustrative example, the steps of a shift-and-leak attack are as follows: 

\begin{enumerate}[leftmargin=*]
\item Identify leaky cells ($LC$s) that can leak info onto a PO.

\item Insert a stuck-at-fault at the chosen $LC$ candidate. 

\item Propagate the fault onto a PO (SCs set to unknown $X's$). If it fails to propagate, it rules out this $LC$, and repeats steps 1 and 2. 

\item Power up the chip in mode $M_0$ to load the correct key into $SC$s. 

\item Switch to mode $M_{1a}$ ($SC$s hold value) and shift in $d$-bit reverse-shifted of the leak condition into the scan. The value of $d$ is the scan distance between the targeted $SC$ and the chosen $LC$.

\item Switch to mode $M_2$ ($SC$s are in the scan), and perform $d$-bit shift to have the leak condition in place and the key in chosen $LC$.

\item Clocklessly switch to mode $M_0$ and observe the PO, to leak the content of the $LC$, i.e., the target key bit.

\end{enumerate}

The authors noted that when the number of $SC$s increases, ATPG may fail to find a leak condition for the chosen $LC$. To address this challenge, by exploiting the conventional SAT attack \cite{subramanyan2015evaluating}, a pre-processing step was added to the shift-and-leak attack, in which the logic cone was treated as a locked combinational circuit considering $RC$s as the primary inputs and $SC$s as the key inputs. The pre-processing phase (which resembles the steps of the conventional SAT attack) is launched as follows: 

\begin{enumerate}[leftmargin=*]
    \item Extract the combinational fan-in cones of the PO.
    \item Obtain a Discriminating Input (DIP) from the SAT tool on the extracted circuit.
    \item Power on the IC in mode $M_0$ ($SC$s capture the actual key).
    \item Switch to $M_{1a}$ ($SC$s hold their values), and shift in the obtained DIP from the SAT tool to the $RC$s.
    \item Clocklessly switch to mode $M_0$ and observe the PO ($eval$ of the SAT attack). Then, go to step 2 until no more DIP found.
\end{enumerate}

\subsection{mR-DFS: Resisting Shift-and-Leak} 

\begin{figure}[t]
    \centering
    \includegraphics[width=\columnwidth]{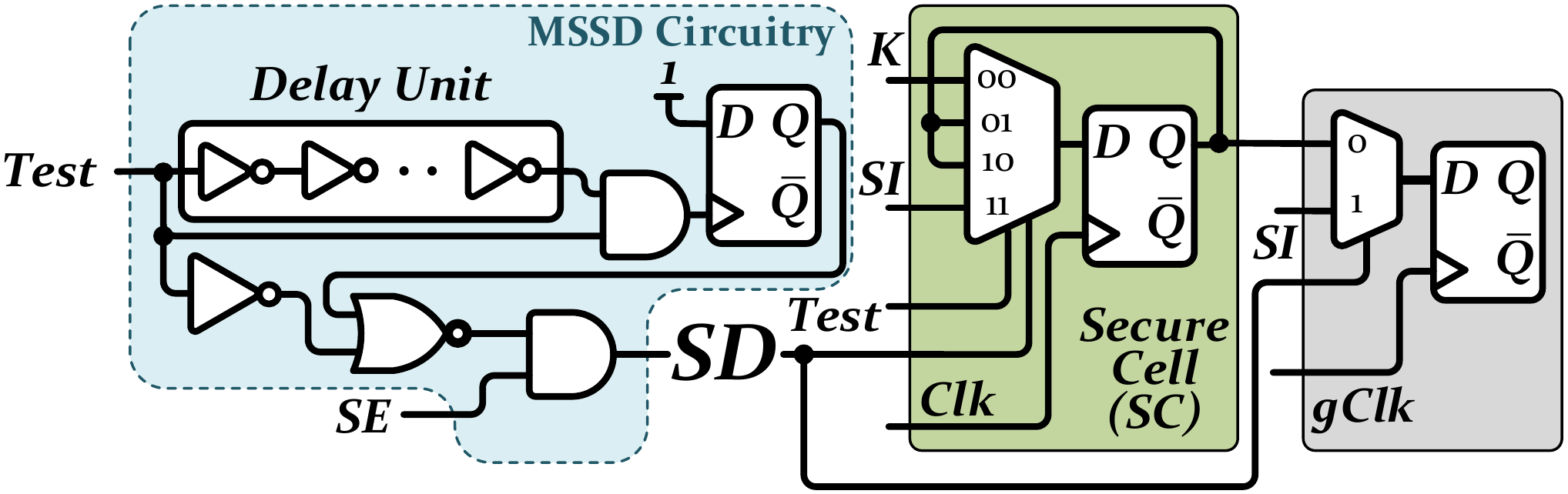}
    \caption{Mode Switch Shift Disable (MSSD) in mR-DFS.}
    \label{mssd}
\end{figure}

As a countermeasure to the shift-and-leak attack, the work in \cite{limaye2019robust} proposes a modified version of robust design-for-security architecture (denoted as mR-DFS in this paper) with a slight modification to the R-DFS. Since mode $M_{1a}$ is used in the shift-and-leak attack to shift-in the known patterns (leak condition or DIP) to $RC$s, in mR-DFS, this mode is blocked. Also, to avoid any other form of leakage, after switching to mode $M_0$, it is not possible to re-enable any shift mode in the scan chain. To do that, as shown in Fig. \ref{mssd}, they build a shift disable ($SD$) signal, such that when $Test$ = 1, $SD$ follows $SE$. But, after the first capture of the actual key, i.e. when the $Test$ is low or when there is a positive transition on the $Test$, $SD$ becomes \emph{ALWAYS ZERO}, thereby blocking the shift operation. Hence, there is no longer a mode where $SC$s can be bypassed, retaining their values, while $RC$s can be loaded/shifted.

\section{\MakeLowercase{m}R-DFS Architectural Drawbacks} \label{drawbacks}

Although mR-DFS addresses the leakage problem in R-DFS using shift disable ($SD$), the introduction of this shift disable ($SD$) signal in mR-DFS poses some new challenges for design and implementation flow, as well as test and debug process. These challenges are discussed next:  

\subsection{High Functional Test Time} \label{high_test}

Since there is no longer mode $M_{1a}$ in mR-DFS architecture, the tester has to rely on mode $M_2$ to shift in and load the RCs. Also, since the shift is disabled when $Test$=0 or after the first positive transition on the $Test$, $Test$ must be high during power ON. Hence, the tester should use $M_2$ as the initial mode to shift in and load the initial state into the RCs. After loading the initial state, the tester switches the mode to $M_0$ to load the actual key. Since it is not possible to re-enable the shift process after switching to mode $M_0$, the tester has to rely on the responses on PO. For the next test pattern, the tester needs to switch back to the mode $M_2$ to shift in and re-load the initial state corresponded to the next pattern. However, due to the blockage of the shift operation after switching to mode $M_0$ ($Test$ = 0), the tester cannot use shift-in anymore for shifting in the next initial state. Hence, the tester has to reset the FF of MSSD circuitry to re-enable shift-in. This reset re-enables the $SD$ to follow $SE$, thereby, the tester can shift in the next initial state. However this reset (\emph{sys\_rst}) will clear all storage elements, including SCs. So, it forces the tester to re-load the keys for the next test pattern. Hence, the actual key must be loaded again from tpNVM to accomplish the functional test, and this key reloading process (with each test pattern) significantly increases the functional test time. It should be noted that the initial state could be chosen to be used for a group of test patterns; however, choosing a specific initial state to be used for a group of patterns would increase the complexity of the functional test significantly. Besides, the designer cannot separate the reset pins for MSSD. Assuming that this reset pin is separated, the adversary can engage it to re-enable shift operation while the actual key is in place.

\subsection{Necessity of Duplicating the SCs} \label{redundantsc}

In mR-DFS, after shifting in the initial state to the RCs using mode $M_2$, the tester switches to mode $M_0$ for only one cycle to load the actual key. However, during this one cycle, the RCs (loaded by initial state) would be updated. To avoid this problem, a clock gating circuitry has been introduced in mR-DFS to disable the clock for one clock cycle after switching from $M_2$ to $M_0$. 

Without any consideration for this requirement in mR-DFS architecture, there are two possible methods to load the actual key from tpNVM in one clock cycle, however, both of them incurs considerable performance/area overhead: (1) engaging an ultra-wide memory that provides all bits of logic obfuscation key at once using only one read operation, (2) engaging temporary registers (FFs) to load the key into them at power ON, then connecting each SC to its corresponding temporary register to be loaded in one clock cycle. 

Regarding the former solution, it is required to have direct wiring from tpNVM to each SC (per each key gate). Hence, the ultra-wide memory must have an extremely high fanout to provide this direct connection. This ultra-high fanout wiring increases the complexity of placement and routing (PnR) process, and it would significantly decrease the performance of the design, and due to optimization constraints in each design, using this scheme is almost impractical.


By choosing the latter method, the incurred overhead is more reasonable. However, the required reset (\emph{sys\_rst}) for loading the next initial state will clear whole registers in the chip. So, a key re-loading from tpNVM to the temporary register is required for each (group of) test pattern. It raises two big problems in mR-DFS: (1) It significantly increases the required time for functional test, and (2) Since key re-loading takes more than one clock cycle, it violates the assumption of mR-DFS, where clock-gating disables the clock signal only for one clock cycle to preserve the value of the RCs. So, after only one clock cycle, during the key re-loading, the RCs would be updated, and the functional test will fail.


\subsection{Re-enabling Shift using Leaky Glitches}
\label{sec:glitch_mr_dfs}

\begin{figure}
    \centering
    \subfloat[MSSD (Blockage) Circuitry in mR-DFS.] {{\includegraphics[width=\columnwidth]{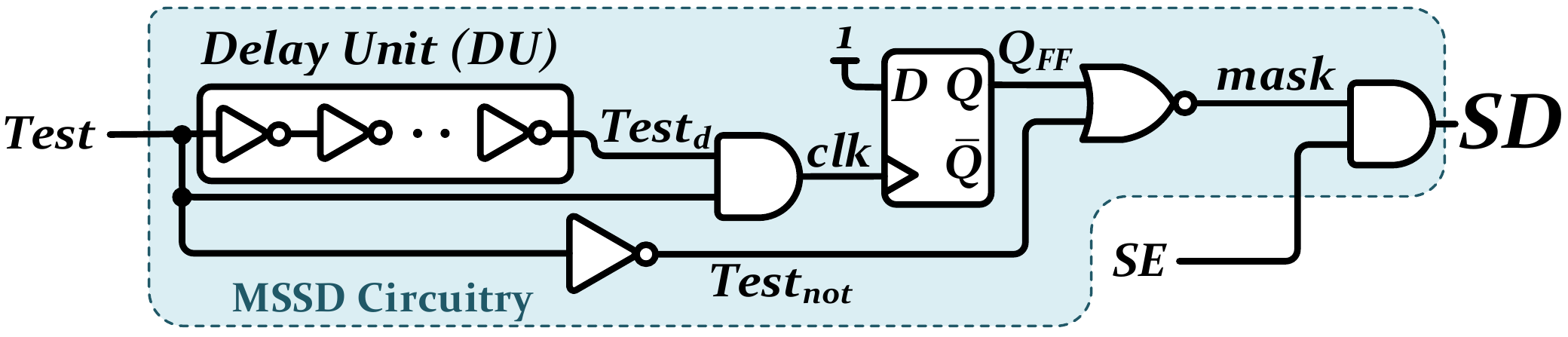}}} \\
    \subfloat[Glitches in Post-Synthesis Timing Simulation of MSSD Circuitry.] {{\includegraphics[width=\columnwidth]{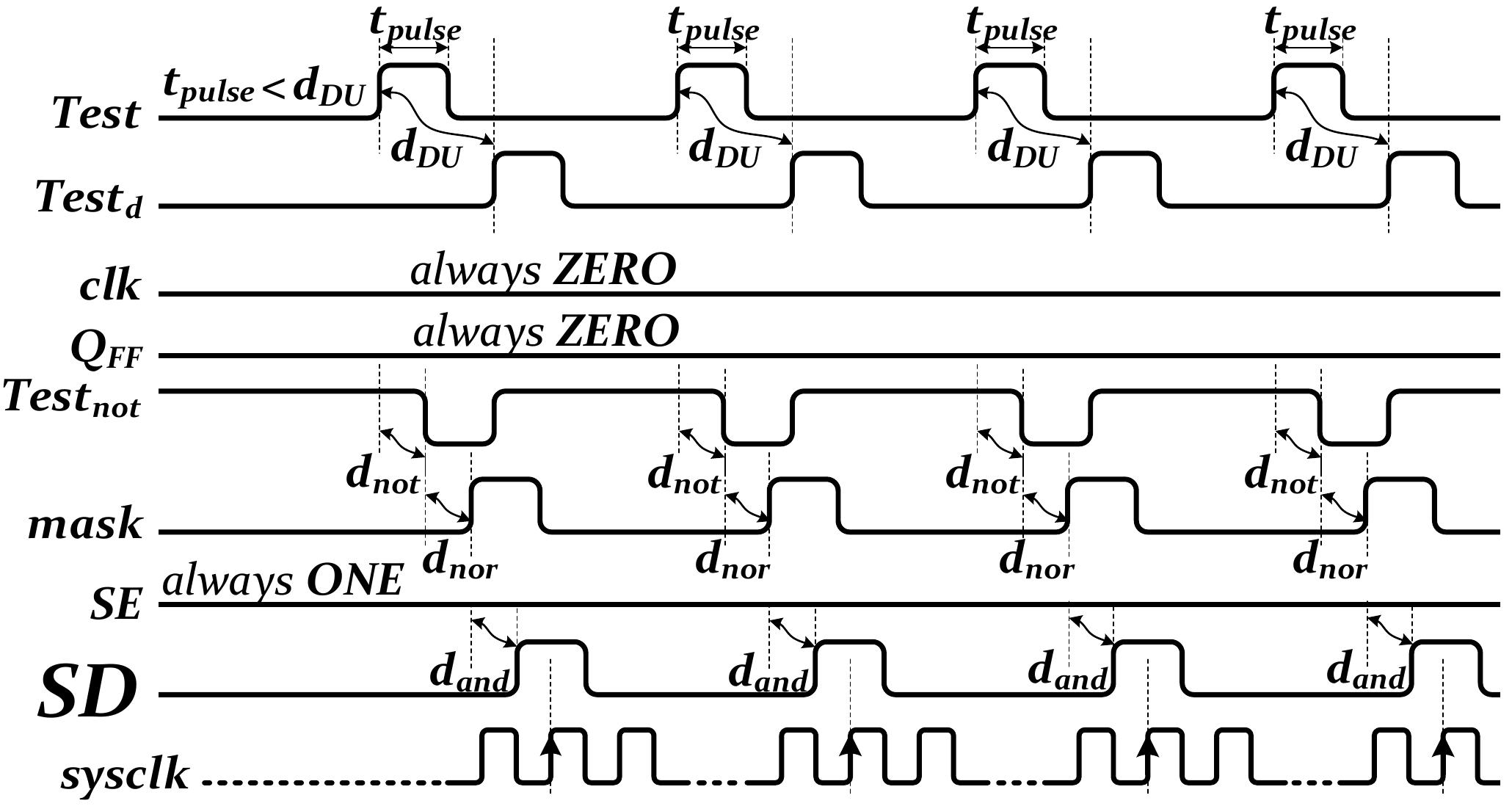}}} \\
    \caption{Re-enabling Shift after Actual Key Load.}
    \label{glitch_a}
\end{figure}

In mR-DFS, as shown in Fig. \ref{mssd}, the selector of $MUX21$ in RCs is controlled by $SD$, which becomes \emph{ALWAYS ZERO} immediately after the first attempt of switching back from mode $M_0$ to $M_2$ (re-enabling shift process). Switching from mode $M_0$ to $M_2$ means that there is a positive transition on the $Test$ pin, and this positive transition allows the FF in MSSD circuitry to capture its input (\emph{CONSTANT ONE}). However, there is still a possibility to switch back from mode $M_0$ to $M_2$ (positive transition on the $Test$ pin) while the FF does not capture its input (\emph{CONSTANT ONE}) to make $SD$ to be \emph{ALWAYS ZERO}. To show that, we draw a timing diagram of the post-synthesis timing simulation of all internal wires of MSSD circuitry. 

As shown in Fig. \ref{glitch_a}(a), a delay unit ($DU$) has been used as a part of the fan-in-cone of the FF in MSSD circuitry, which is built using 10 inverters \cite{limaye2019robust}. Assuming that the adversary is aware of timing information of the circuit, as shown in Fig. \ref{glitch_a}(b), she generates a stimuli for $Test$ pin in which the duration of high pulses is less than the delay of $DU$ ($t_{pulse} < d_{DU}$). Hence, the inputs of the first $AND$ gate, i.e. $Test$ and $Test_d$, have no overlap when both signals are high, and accordingly, DFF's $clk$ would be \emph{ALWAYS ZERO}. Since it is assumed that the DFF sets to 0 on reset, $Q_{FF}$ would also be \emph{ALWAYS ZERO}. So, the function of $NOR$ gate is similar to $NOT$ gate, whose input is $Test_{not}$. Consequently, $mask$ follows $Test$ with a delay of $d_{not} + d_{nor}$, and similarly, if we suppose that $SE$ is \emph{ALWAYS ONE}, $SD$ follows $Test$ with a delay of $d_{not} + d_{nor} + d_{and}$. Since $SD$ controls the shift operation in mR-DFS, using these potential glitches, the $SD$ can re-enable the shift operation after mode $M_0$. 


\section{Proposed Solution}
\label{sec:proposed}

When the logic obfuscation is in place, to introduce a secure and robust scan chain architecture, three requirements must be met:

\begin{enumerate}[leftmargin=*]
    \item There must be no possibility of key leakage during the test.
    \item Both structural test and functional test must be carried out in a reasonable time (low test time overhead compared to the test time of the original design) without significant loss of coverage. 
    \item The complexity of test flow (structural and functional) and the overhead of secure scan chain architecture must be minimized.
\end{enumerate}

\begin{figure}
    \centering
    \subfloat[Decoupling SCs from RCs] {{\includegraphics[width=0.525\columnwidth]{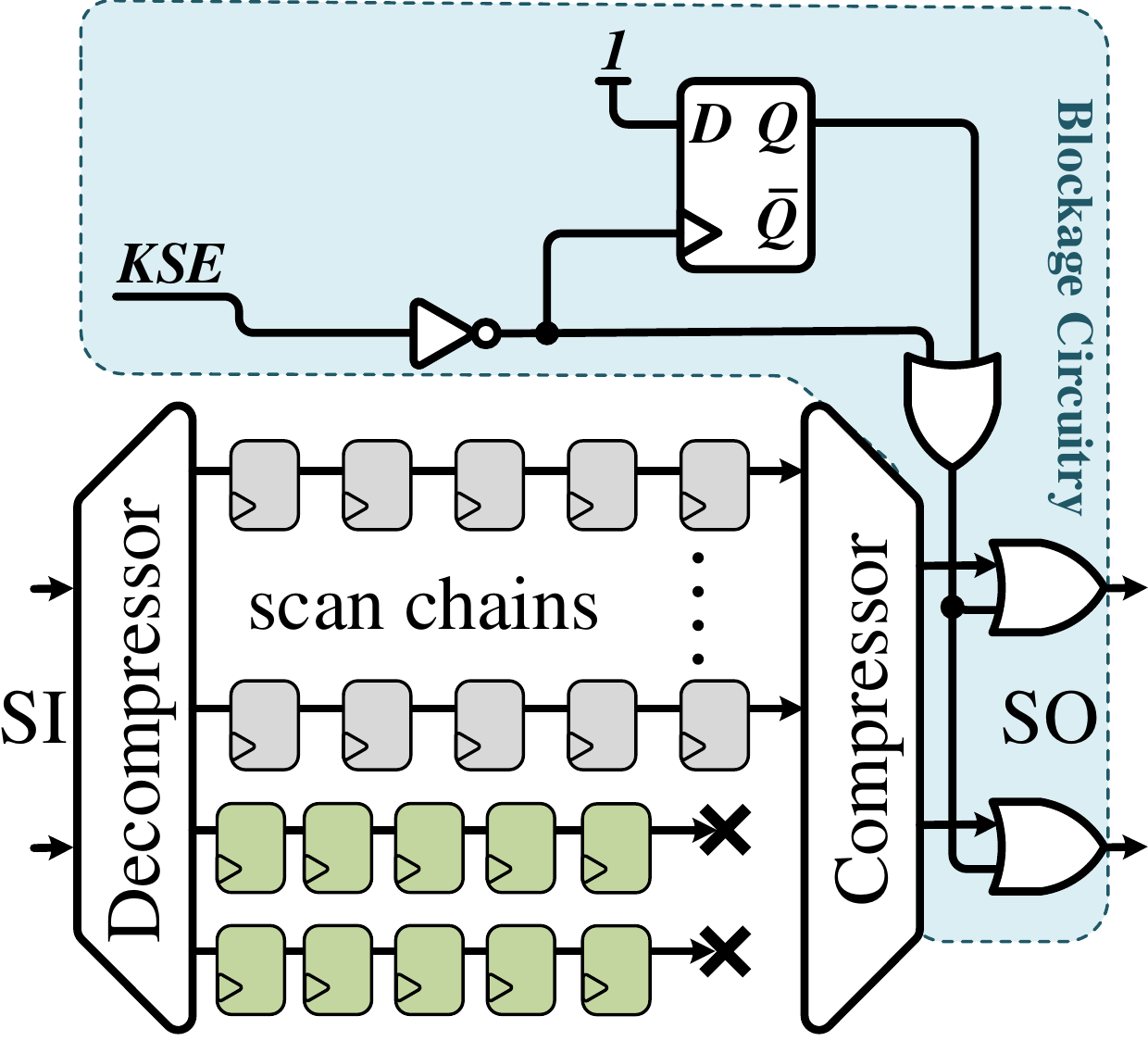}}}
    \subfloat[One-way Secure Cell (1wSC)] {{\includegraphics[width=0.475\columnwidth]{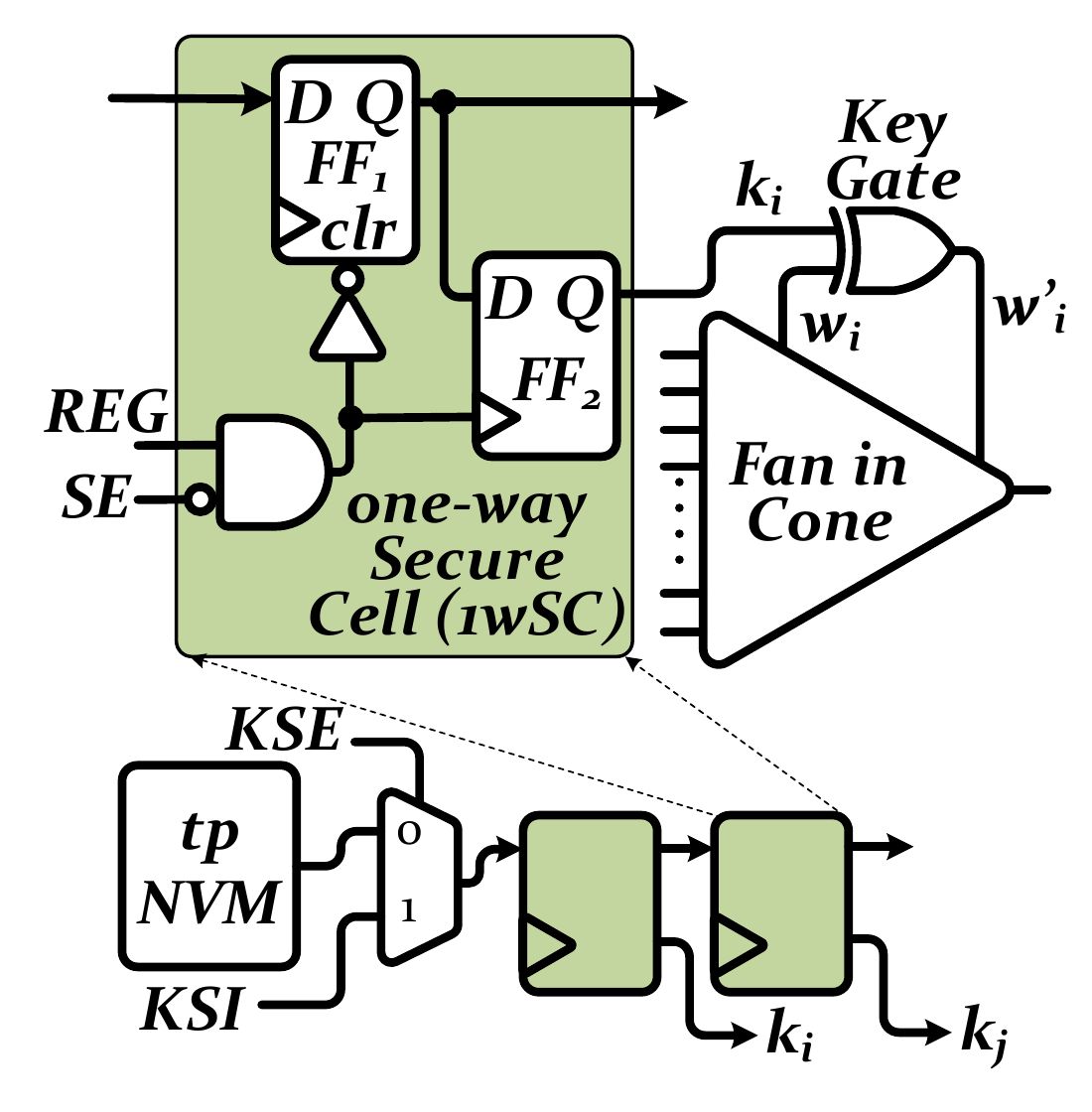}}}
    \caption{(a) kt-DFS Architecture with New Re-designed Blockage Circuitry, (b) Using 1wSC for Logic Locking Key. $KSE$ determines the source of the key (tpNVM or $KSI$).}
    \label{sc_irdfs}
\end{figure}

In our proposed key-trapped DFS (kt-DFS), the scan chain(s) of the SCs are completely decoupled from the scan chain(s) of the RCs. In fact, \textbf{{there is no reason for stitching the RC and SC cells in one chain, which has been the source of vulnerability in both R-DFS and mR-DFS}}. As illustrated in Fig. \ref{sc_irdfs}(a), there is no common path between RCs and SCs in our proposed kt-DFS architecture. Also, considering that the SCs are only responsible to store the key value, none of the internal operations/computations overwrites the content of the SCs. So, when the scan chain is in place for the SCs, only the shift-in through SI is available for them to load the keys, and the SO is permanently blocked for scan chain(s) of the SCs. 

To guarantee the security of SCs against any form of leakage, we re-design and introduce a new secure cell, called 1-way secure cell (1wSC). Fig. \ref{sc_irdfs}(b) depicts the details of 1wSC. Each 1wSC has two internal storage elements: a scan-connected storage (denoted as $FF_1$), and a trap storage (denoted as $FF_2$). The scan-connected storage could be used to shift values in and out of the 1wSC or into the trap storage. However, the value of the trap storage cannot be shifted out, and is only connected to its corresponded key gate. The transfer of key value from $FF_1$ to $FF_2$ takes place after setting $REG = 1$ and $SE = 0$, which is called \emph{register mode}. Registration of the key into trap storage takes place on the rising edge of the clock input of the $FF_2$, which is a function of $REG$ and $SE$. Also, this condition is used as the $RESET$ condition of all $FF_1$s to clear their values. Hence, $AND(Test, \overline{SE})$ is used as the clock source of $FF_2$s, and its toggled is used as the $RESET$ for $FF_1$s. 

Also, the trap storage does not have a reset, and upon power-up randomly initialized to 0 or 1. So, upon transition of the key from scan-connected storage to the trap storage, since the storage is initialized randomly, the adversary cannot determine the previous value of the trap storage. This prevents the back-side imaging attack based on the captured heat map as described in \cite{rahman2020key} (e.g. when the activity is observed on heat map for a specific storage element, the adversary cannot determine if the transition is \{0 $\rightarrow$ 1\} or \{1 $\rightarrow$ 0\}, and if \emph{NO} activity is observed, the adversary cannot determine if the transition is \{0 $\rightarrow$ 1\} or \{1 $\rightarrow$ 1\}).

In our proposed kt-DFS, the keys could be loaded into 1wSC from either tpNVM or scan-in (SI). Hence, the tester would be able to carry out the structural test by loading the desired key using SI. But, since the scan chain(s) of 1wSCs are decoupled in kt-DFS, two dedicated scan-enable and scan-in are used for the scan chain(s) of the SCs, called $KSE$ and $KSI$ respectively.

The behavior of 1wSC is controlled using two pins, here called $REG$ and $SE$. As captured in Table \ref{new_mode}, based on these two pins, a 1wSC can be operated in three main modes: 

\begin{enumerate}[leftmargin=*]
\item \textbf{Functional Mode ($M_0$)}: \{$REG$, $SE$\}=\{0,0\}, and the RCs are in capture mode. Trap storage (FF$_2$s) must have the key. However, scan-connected storage (FF$_1$s) is able to capture a new key.  
\item \textbf{Shift Mode ($M_{1,3}$)}: \{$REG$, $SE$\}=\{$X$,1\}, and the RCs are in shift mode. Scan-connected storage (FF$_1$s) is able to capture the key simultaneously, and there is no action on trap storage (FF$_2$s).
\item \textbf{Register Mode ($M_2$)}: \{$REG$, $SE$\}=\{1,0\}, and the pre-loaded key in scan-connected storage (FF$_1$s) would be written to trap storage (FF$_2$s), and scan-connected storage (FF$_1$s) will be cleared. 
\end{enumerate}


Similar to R-DFS and mR-DFS, a blockage circuitry is required to block the SO after the first attempt of key loading from the tpNVM. To support our proposed operational modes in the kt-DFS, a new blockage circuitry is designed. In kt-DFS, the SO must be blocked after loading the actual keys into $FF_2$s. When $KSE$ is low, the $FF_1$ is fed using tpNVM. Hence, $\overline{KSE}$ is used to mask the SO. Note that the actual key would be loaded into $FF_2$ when $REG$ = 1 and $SE$ = 0 (register mode). However, before this condition, the tester has to load the actual key into $FF_1$s while the $KSE$ is low. Hence, by only considering $KSE$ = 0 as the blocking condition, we also cover the register-mode. Accordingly, the SO would be no longer available when $KSE$ becomes low. 

\begin{table}
\footnotesize
\centering
\caption{Modes of Operation in kt-DFS.}
\label{new_mode}
\setlength\tabcolsep{4pt} 
\begin{tabular}{@{} cccl @{}}
\toprule 
REG & SE & Mode & Description \\
\cmidrule(lr){1-2} \cmidrule(lr){3-3} \cmidrule(lr){4-4}
\multirow{2}{*}{0} & \multirow{2}{*}{0} & $M_0$ & $FF_2$ must have the key*.  \\
 & & (Functional Mode) & $FF_1$ could capture the key*. \\
\cmidrule(lr){1-2} \cmidrule(lr){3-3} \cmidrule(lr){4-4}
\multirow{2}{*}{0} & \multirow{2}{*}{1} & $M_1$ & \multirow{2}{*}{$FF_1$ could capture the key*.}  \\ 
 & & (Shift Mode) & \\ 
\cmidrule(lr){1-2} \cmidrule(lr){3-3} \cmidrule(lr){4-4}
\multirow{2}{*}{1} & \multirow{2}{*}{0} & $M_2$ & $FF_2$ are fed from $FF_1$.  \\ 
 & & (Register Mode) & $FF_1$ will be reset to \emph{ZERO}. Chain is erased. \\
\cmidrule(lr){1-2} \cmidrule(lr){3-3} \cmidrule(lr){4-4}
\multirow{2}{*}{1} & \multirow{2}{*}{1} & $M_3$ & \multirow{2}{*}{$FF_1$ could capture the key*.}  \\ 
 & & (Shift Mode) & \\ 
\bottomrule
\multicolumn{4}{l}{* Based on $KSE$, actual/dummy key could be loaded from tpNVM/$KSI$}
\end{tabular}
\end{table}

\subsection{No Possibility of Key Leakage in kt-DFS}

Considering that the leakage problem in R-DFS and mR-DFS is for unnecessary stitching of the RC and SC in the same scan chain, we fully decouple the SCs and RCs scan chains in kt-DFS, and the output of the scan chain(s) of SCs is permanently blocked. The values stored in the scan-connected storage ($FF_1$s) will be cleared with the transfer of the key to the trap storage ($FF_2$s). This guarantees that key values are trapped and no either regular or glitch-based shift can leak the key values to the SO.

\subsection{Functional/Structural Test in kt-DFS}

In kt-DFS architecture, the functional test and the structural test could be done without any significant limitation or any substantial overhead. For the structural test, since the SCs are equipped with new $KSE$ and $KSI$ pins, it could be accomplished using the following steps:  

\begin{enumerate}[leftmargin=*]
    \item Set $KSE \rightarrow$ 1 and mode to $M_0$. Shift in a dummy key via $KSI$. 
    \item Switch to mode $M_2$ to write the key into $FF_2$, and to clear $FF_1$. 
    \item Switch to mode $M_1$ to shift in the initial state into RCs. 
    \item Switch to mode $M_0$ for one clock cycle for capturing new state. 
    \item Switch again to mode $M_1$ to shift out the RCs to SO. 
\end{enumerate}

Unlike the structural test, the functional test requires the actual key. Hence, loading the key from tpNVM followed by register mode will block the SO. Considering the blockage of the SO, the steps of the functional test is as follows: 

\begin{enumerate}[leftmargin=*]
    \item Set $KSE \rightarrow$ 0 and mode to $M_0$. Shift in the actual key from tpNVM. (When $KSE$ = 0, the SO is blocked.)
    \item Switch to mode $M_2$ to write the key into $FF_2$, and to clear $FF_1$. (Once $KSE$ = 0 and mode is $M_2$, the SO will no longer available.) 
    \item Switch to mode $M_1$ to shift in the initial state into the RCs. 
    \item Switch to mode $M_0$ for one clock cycle for capturing new state, and clocklessly observe the PO.
\end{enumerate}

It should be noted that similar to R-DFS and mR-DFS, the tester accomplishes the functional test through observing the PO with negligible loss of coverage. 

\subsection{Test Complexity and Scan Chain Overhead}

\begin{table}
\footnotesize
\centering
\caption{Area Overhead, Test Coverage, and Leakage Comparison between R-DFS, mR-DFS using MSSD, and proposed kt-DFS for Identical Timing Constraints. (Key Size = 128, Number of Scan Chain = 1)}
\label{gen_res}
\setlength\tabcolsep{4pt} 
\begin{tabular}{@{} c c ccc @{}}
\toprule 
\multirow{3}{*}{Benchmark} & Original & \multicolumn{3}{c}{R-DFS \cite{guin2018robust}} \\
\cmidrule(lr){2-2} \cmidrule(lr){3-5} 
 & Test Coverage  & Area Overhead & Test Coverage & Key Recovered \\  
 & (\%) & (\%) & (\%) & (\#) \\
\cmidrule(lr){1-1} \cmidrule(lr){2-2} \cmidrule(lr){3-5}
s35932 & 100 & 12.49\% & 100 & 127 \\
s38417 & 100 & 13.84\% & 100 & 128 \\
s38584 & 100 & 14.85\% & 100 & 128 \\ 
\cmidrule(lr){1-1} \cmidrule(lr){2-2} \cmidrule(lr){3-5} 
b17 & 99.91 & 6.24\% & 99.72 & 127 \\
b18 & 99.77 & 3.08\% & 99.78 & 126 \\
b19 & 99.8 & 1.17\% & 99.78 & 127 \\
\midrule
\midrule
\multirow{3}{*}{Benchmark} & Original & \multicolumn{3}{c}{mR-DFS using MSSD \cite{limaye2019robust}} \\
\cmidrule(lr){2-2} \cmidrule(lr){3-5} 
 & Test Coverage  & Area Overhead & Test Coverage & Key Recovered \\  
 & (\%) & (\%) & (\%) & (\#) \\
\cmidrule(lr){1-1} \cmidrule(lr){2-2} \cmidrule(lr){3-5} 
s35932 & 100 & 8.16\% & 100 & 127 \\
s38417 & 100 & 9.31\% & 100 & 128 \\
s38584 & 100 & 11.17\% & 100 & 128 \\ 
\cmidrule(lr){1-1} \cmidrule(lr){2-2} \cmidrule(lr){3-5}
b17 & 99.91 & 4.79\% & 99.69 & 127 \\
b18 & 99.77 & 1.75\% & 99.77 & 126 \\
b19 & 99.81 & 0.67\% & 99.79 & 127 \\
\midrule
\midrule
\multirow{3}{*}{Benchmark} & Original & \multicolumn{3}{c}{Proposed kt-DFS} \\
\cmidrule(lr){2-2} \cmidrule(lr){3-5} 
 & Test Coverage  & Area Overhead & Test Coverage & Key Recovered \\  
 & (\%) & (\%) & (\%) & (\#) \\
\cmidrule(lr){1-1} \cmidrule(lr){2-2} \cmidrule(lr){3-5} 
s35932 & 100 & 5.21\% & 100 & 0 \\
s38417 & 100 & 5.91\% & 100 & 0 \\
s38584 & 100 & 6.27\% & 100 & 0 \\ 
\cmidrule(lr){1-1} \cmidrule(lr){2-2} \cmidrule(lr){3-5}
b17 & 99.91 & 1.84\% & 99.67 & 0 \\
b18 & 99.77 & 0.55\% & 99.73 & 0 \\
b19 & 99.8 & 0.24\% & 99.78 & 0 \\
\bottomrule
\end{tabular}
\end{table}

Decoupling the scan chain(s) of SCs from that of RCs helps to facilitate the test flow for the tester compared to the test flow in mR-DFS. Despite mR-DFS with a mandatory \emph{sys\_rst} for each (group of) test pattern, no additional operation is required in kt-DFS for any form of the test. No \emph{sys\_rst} is required, and none of the operations is blocked after the first attempt of the actual key loading from tpNVM, and similar to R-DFS, only the SO is blocked to break the SAT attack. However, unlike R-DFS, it is fully secure against any form of leakage-based attacks, such as shift-and-leak.  

The 1wSC in our proposed kt-DFS has two storage units and has a larger footprint compared to the SCs used in R-DFS and mR-DFS. However, the R-DFS and mR-DFS also need to transfer the key values from tpNVM to SCs. Using a very wide memory to derive thousands of keys is quite demanding in terms of area, and it imposes higher complexity during PnR. Hence, the R-DFS and mR-DFS also need to resort to a chain of temporary registers to transfer the keys. This means there is also a duplicated register per each SC in both R-DFS and mR-DFS. Furthermore, compared to $MUX41$ in both R-DFS and mR-DFS, only one $AND$ gate and one $NOT$ have been used in each 1wSC, which slightly improves the area overhead.


\section{Experimental Result}
\label{sec:Result}

To analyze the security of the kt-DFS, and to provide better comparative results, we engage the same ITC-99 and ISCAS-89 benchmark circuits as used in mR-DFS \cite{limaye2019robust}. We engaged strong logic locking (SLL) \cite{rajendran2012security} in all experiments to determine the location of key gates, and the number of key bits is 128. All the experiments have been accomplished on a Dell PowerEdge R620 equipped with Intel Xeon E5-2670 2.50GHz and 64GB of RAM, using Synopsys Design Compiler 2017.09, Tetramax 2017.09, and VCS 2017.12 tools along with the Synopsys generic 32nm library.

Table \ref{gen_res} represents the area overhead, test coverage, and the leakage of R-DFS, mR-DFS, and our proposed kt-DFS. The number of (regular) scan chains (composed of RCs) is set to be one. This is because with a large number of scan chains, the length of each scan chain will be short. Hence, the chosen LCs can leak the content of a smaller number of SCs from their shorter scan chains, which decreases the success rate of the shift-and-leak attack. Hence, we assume that there is only one scan chain in the circuit to make it the best-case scenario for the shift-and-leak.

Considering that the access to the scan chain is restricted, the SAT attack cannot be deployed. This does not prevent an attacker from deploying the unrolling or bounded-model-checking (BMC) \cite{el2017reverse} attack that only relies on PI/PO. However, this group of attacks runs into scalability issues as they rely on two sub-routines which are in PSPACE and NP \cite{limaye2019robust}. Even the accelerated version of this attack (described in \cite{shamsi2019kc2}) fails to terminate for even small designs. Besides, new techniques such as DFSSD \cite{roshanisefat2020dfssd} shows how low overhead techniques, like deep faults and shallow state duality, could be used to break the state-of-the-art sequential SAT attacks. Hence, Table \ref{gen_res} only reflects the effectiveness of the shift-and-leak attack. 

To evaluate the possibility of the leakage in R-DFS, we engage the shift-and-leak attack in \cite{limaye2019robust} with no change. However, for mR-DFS, we integrated the glitch propagation that described in Section \ref{sec:glitch_mr_dfs} with the shift-and-leak attack. Although the mR-DFS is resilient against the original shift-and-leak attack, assuming the potential glitches in MSSD, as well as controlling the \emph{sys\_clk} using an external clock generator, the security of mR-DFS and R-DFS is at the same level, and both could be broken by leaking the key bits onto PO. However, there is no leakage possibility in the proposed kt-DFS, and the adversary cannot recover the content of any SC. 

Regarding the area overhead, we assume that both R-DFS and mR-DFS use temporary registers to transfer the key values from tpNVM to SCs (this has lower overhead compared to using a very wide memory). Hence, as a part of the logic obfuscation circuit, these temporary registers affect the area overhead. Note that in our proposed kt-DFS, these temporary registers are part of 1wSC. Also, compared to SCs in R-DFS and mR-DFS less basic gates are used in kt-DFS. Overall, compared to R-DFS and mR-DFS, kt-DFS reduces the area overhead by 61\% and 44\%, respectively.

As discussed previously, all three schemes block the SO after the first attempt of key loading from tpNVM. Hence, all have to rely on the PO for the functional test. As shown in Table \ref{gen_res}, the test coverage loss is negligible and almost identical in all schemes.

\section{Conclusion}
\label{sec:Conclusion}

In this paper, we first evaluated the information/key leakage possibility and design methodology drawbacks of recently published DFS architectures in the presence of scan chain locking. Then, we proposed a new obfuscated DFS solution, denoted as key-trapped DFS (kt-DFS) that addresses the prior art shortcomings. In kt-DFS, we introduced a new secure storage cell for the storage of key values. The proposed secure cell allows us to trap the key after being loaded, preventing different forms of shift and leak attacks (glitch based or logic-based), while safely removing the key upon transition from functional to test mode. At the same time, we illustrated that using the proposed DFS, the design can safely undergo manufacturing and functional testing without incurring any significant limitation in terms of increase in the test time (functional or manufacturing) while maintaining desirably low overhead. 

\bibliographystyle{splncs04}
\bibliography{refs}

\end{document}